\documentclass[review]{elsarticle}

\usepackage{graphics}
\usepackage{graphicx}
\usepackage{float}%place images here with [H]
\usepackage{bm}
\usepackage[nomarkers,notablist,nofiglist]{endfloat}
\usepackage{lineno,hyperref}
\modulolinenumbers[5]
\usepackage{siunitx}         
\graphicspath{ {./Figures/} } 
\usepackage{soul}

\usepackage{geometry} % to change the page dimensions
\usepackage{color} 
\usepackage{outlines}
\journal{arXiv}

\newcommand\blfootnote[1]{%
  \begingroup
  \renewcommand\thefootnote{}\footnote{#1}%
  \addtocounter{footnote}{-1}%
  \endgroup
}

\begin{document}
\begin{frontmatter}
\title{Analysis of a Three-Dimensional Slip Field in a Hexagonal Ti Alloy from in-situ High-Energy X-ray Diffraction Microscopy Data}
\author{Darren C. Pagan\textsuperscript{1*}\blfootnote{*Corresponding Author}}
\author{Kelly E. Nygren\textsuperscript{2}}
\author{Matthew P. Miller\textsuperscript{2,3}}
\address{\textsuperscript{1}The Pennsylvania State University, University Park, PA 16802 USA}
\address{\textsuperscript{2}Cornell High Energy Synchrotron Source, Ithaca, NY 14853 USA}
\address{\textsuperscript{3}Cornell University, Ithaca, NY 14853 USA}

\begin{abstract}
Here we analyze a three-dimensional distribution of crystallographic slip measured in-situ during the uniaxial deformation of hexagonal Ti-7Al. The slip field is reconstructed using a novel methodology that combines spatially resolved lattice orientation fields and grain-averaged stresses measured using high-energy X-ray diffraction microscopy (HEDM) with crystal plasticity. Analysis is performed to explore lattice orientation dependence, stress dependence, and connectivity (network relationships) of grains experiencing elevated amounts of slip. Elevated slip is found to be primarily associated with a single large network of connected grains, and within this network, a clustered group of grains oriented favorably for slip are found to have outsized structural importance. The effect of different rate sensitivities of families of slip systems on reconstructed slip activity is also discussed.
\end{abstract}

\end{frontmatter}

\section{Introduction}

Our experimental understanding of how crystallographic slip (or simply slip) proceeds in polycrystalline engineering alloys has primarily been driven by observations at or near the sample surface. While numerous failure processes are often dominated by the growth of surface defects mediated by slip, other ductile failure processes such as void coalescence and shear band formation are believed to be initiated in the bulk of a deforming polycrystal due to heterogeneity of slip at subgrain length scales. These slip-driven failure processes are inherently three-dimensional (3D). Fully understanding these failure processes creates a need to probe how slip is transferred to and around possible nucleation points \emph{in 3D as it is occurring}. To date, measurements of slip fields in the bulk of a polycrystal have not been possible, although this is changing. New synchrotron X-ray techniques have demonstrated that bulk slip characterization is possible in single crystals \cite{pagan2016determining,pagan2018analyzing} and that the stress states driving slip in individual crystals within polycrystals can be determined \cite{wang2017direct,bhattacharyya2021elastoplastic}. In this work, we extend these efforts to analyze full 3D slip fields in deforming polycrystals in-situ. The slip fields are reconstructed from evolving 3D orientation distributions measured using near-field high-energy X-ray diffraction microscopy (nf-HEDM) interpreted through continuum slip kinematics. One traditional challenge for analyzing slip behavior, no knowledge of the local stress states driving slip, is overcome through complementary usage of the far-field variant of high-energy X-ray diffraction microscopy (ff-HEDM).

Characterizing how slip is organized and transferred is particularly important in hexagonal-close-packed (HCP, $\alpha$ phase) titanium alloys. Fatigue and fracture mechanisms have been directly tied to various slip transmission conditions including relatively large volumes of grains containing similar lattice orientation (microtextured regions or macrozones) \cite{pilchak2014simple} and specific pairings of grains that develop stress concentrations (hard and soft grain pairings) \cite{dunne2008mechanisms}. Currently our knowledge regarding these slip processes has been solely derived from computational simulation and post-mortem microscopy. Here we present the first full analysis of 3D slip fields and a network structure of slip in a hexagonal Ti alloy gathered in-situ which can help to validate failure theories and improve our ability to predict the time and location of failure in Ti alloys containing $\alpha$ phase.

As mentioned, traditional methods of slip characterization have been primarily limited to studies of the sample surface or through destructive methods which expose new surfaces that can be analyzed. Efforts include measuring the orientation of slip traces observed on a sample to find the orientation of active slip planes \cite{bieler2014grain,bieler2019analysis,kasemer2017slip}, etching the sample surface to expose termination points of dislocations with the sample surface \cite{kocks1958observations,rosenbaum1961non}, and using low-energy polychromatic X-ray beams to characterize lattice distortion \cite{kocks1964compression,ice2007white}. More recent efforts are capable of characterizing the geometrically necessary dislocations produced by slip with high-resolution electron back scatter diffraction \cite{wilkinson2010high}, strain fields produced around small numbers of slip events with high-resolution digital image correlation \cite{stinville2016sub}, and the orientation and height of slip traces with scanning electron microscopy measurements \cite{bourdin2018measurements}. However, there are two issues with these types of surface measurements which limit the generality of conclusions drawn from them. The first is that the stress state of grains at the surface will vary significantly from those deeper within the material. More importantly, heterogeneous slip caused by interactions between grains in the bulk of a polycrystal are not monitored, leaving the underlying drivers for deformation unknown. Micropillar mechanical testing from which slip behavior can be analyzed in 3D with significantly less ambiguity \cite{uchic2009plasticity,maass2007time} partially addresses some of these issues, but the boundary conditions in micropillar testing do not necessarily reflect those experienced by grains embedded in polycrystals. We note that a common thread among all slip analyses is that they generally consist of measuring changes to the shape of a specimen or the crystal lattice and then inferring from a model the crystallographic slip that must have occurred to produce the change (often while making assumptions about the microscale stress state). The methodology presented in this paper also follows a model-based approach to slip measurement interpretation, but the HEDM probe enables the analysis to occur in the bulk of the polycrystal.

Over the past twenty years, HEDM techniques (also referred to as 3D X-ray Diffraction \cite{poulsen2004three}) performed at synchrotron sources have provided new capabilities to explore microstructural and  micromechanical evolution in engineering alloys as they are thermomechanically loaded \cite{miller2020understanding}. As these techniques have matured, improvements to efficiency of reconstruction algorithms have enabled more grains to be probed simultaneously and improvements in detection speed have increased the number of material states (points on the stress-strain curve) that can be probed during a measurement \cite{oddershede2012measuring, schuren2015new, beaudoin2017bright}. Additionally, advances in sample environments have made multimodal measurements more commonplace \cite{shade2015rotational}. In particular, nf-HEDM, which enables 3D spatial distributions of lattice orientation to be measured non-destructively and ff-HEDM, which allows for full grain-averaged elastic strain states to be measured, are now readily performed in the same experiment. This multi-modal characterization provides new opportunities to combine these data to provide a more complete picture of material evolution during elastic-plastic deformation. In this work, the micromechanical loading conditions driving slip and the measured lattice deformation in 3D are considered simultaneously, overcoming the inability to characterize micromechanical loading conditions which has historically been a weakness of many slip analyses. Commonly, substitutes for local applied stress conditions, such as isostress assumptions combined with Schmid factor analysis, create uncertainties in interpretation of the slip measurements. In this work, inclusion of local stress state is an integral portion of the reconstruction procedure, leading to self-consistency between reconstructed slip and the micromechanical conditions that produce it.

The novel slip reconstruction method presented here is applied to studying the distribution of slip that develops in Titanium 7\% weight Aluminum (Ti-7Al) deformed in uniaxial tension past the elastic-plastic transition. Ti-7Al is a single phase titanium alloy with only the $\alpha$ phase present. With a 3D reconstruction of a miscroscale slip field available, hypotheses regarding the slip behavior are evaluated including those related to lattice orientation, stress, and influence of neighborhood.  In this work, vectors and tensors are indicated with bold text ($\bm{A}$, $\bm{a}$). A tilde above a quantity indicates a macroscopic value ($\tilde{\bm{A}}$), while an overbar ($\bar{\bm{A}}$) indicates a grain-averaged quantity.

\section{Material and Methods}
\label{sec:exp}

\subsection{Material}
The Ti-7Al tested was cast as a 75 mm ingot, hot isostatically pressed and extruded into a bar with a 30 mm$^2$ cross sectional area. The material was recrystallized at 955$^\circ$C for 24 h and furnace cooled. The test specimen was extracted with the tensile axis parallel to the extrusion direction near the center of the extrusion where the average equiaxed grain-size was 110 $\mu$m. The material was then cut into tensile specimens compatible with the RAMS2 loading device \cite{shade2015rotational}. The cross section of the specimen was nominally a 1 mm by 1 mm square. 

\subsection{Experimental Method}
\begin{figure}[h]
	\centering
	\includegraphics[width=0.7\textwidth]{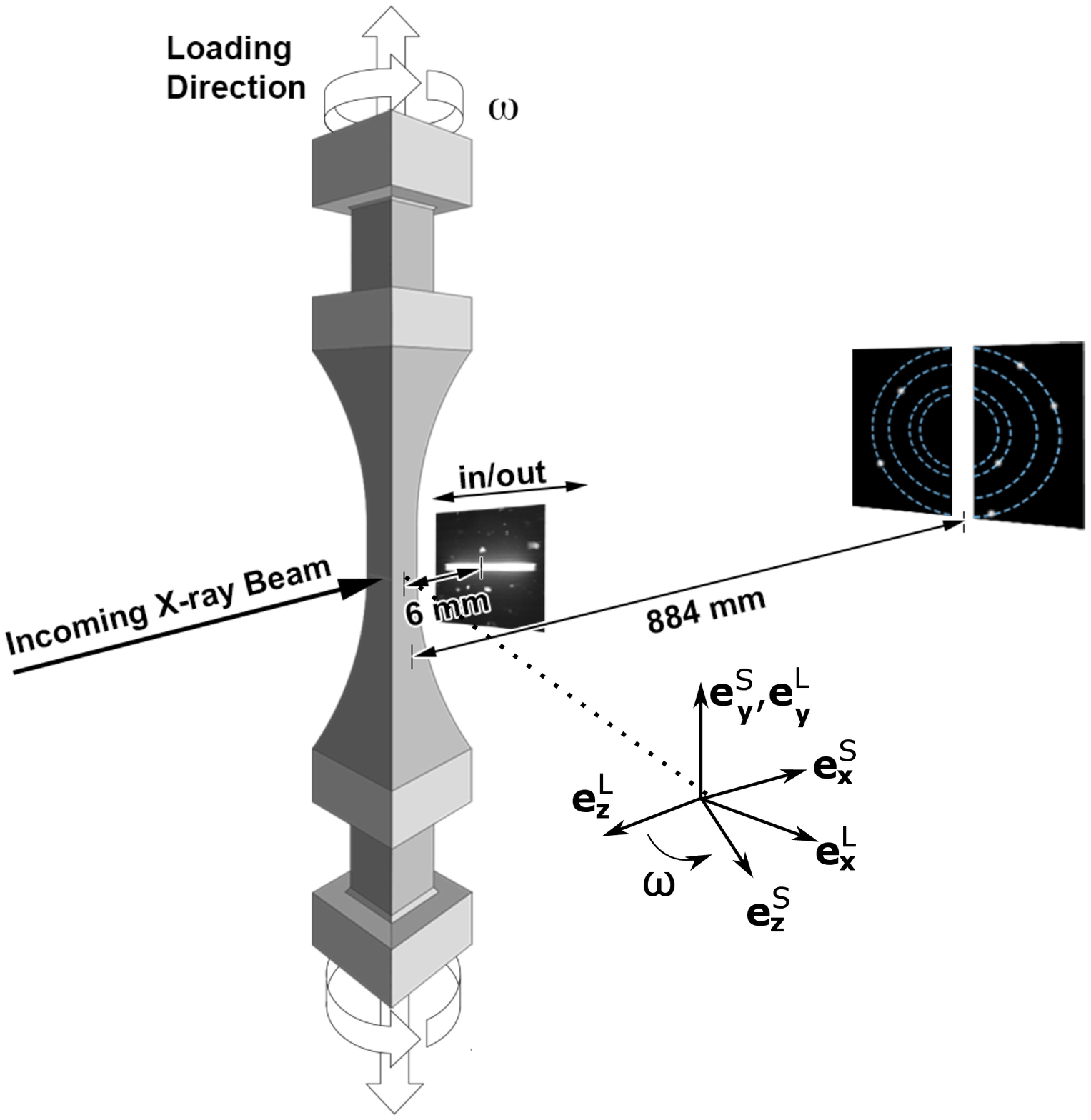}
	  \caption{Schematic of the experimental geometry for high-energy diffraction microscopy (HEDM) measurements. The near-field detector is moved out of the path of diffracted X-rays for far-field measurements. The laboratory coordinate system is denoted with L and the sample coordinate system attached to the rotating tensile specimen is denoted with S. The specimen is loaded along the vertical direction $\bm{e_y}^L$ and also rotated by angle $\omega$ about this direction during X-ray measurements. Locations of the near- and far-field detectors are shown.}
	  \label{fig:geometry}
\end{figure}

The specimen was deformed in uniaxial tension at Beamline ID-1A3 at the Cornell High Energy Synchrotron Source (CHESS). The experimental geometry is illustrated in Fig. \ref{fig:geometry}. A coordinate system associated with the laboratory is labeled L and a second coordinate system fixed to the specimen is labeled S. The incoming beam with an energy of 61.332 keV travels in the $-\bm{e_z}^L$ direction. The specimen was loaded along and rotated around $\bm{e_y}^L$ which is parallel to $\bm{e_y}^S$. The laboratory and specimen coordinate systems are related by a rotation angle $\omega$. We also associate a crystal coordinate system, C, with the lattice of each grain.

\begin{figure}[h]
	\centering
	\includegraphics[width=0.7\textwidth]{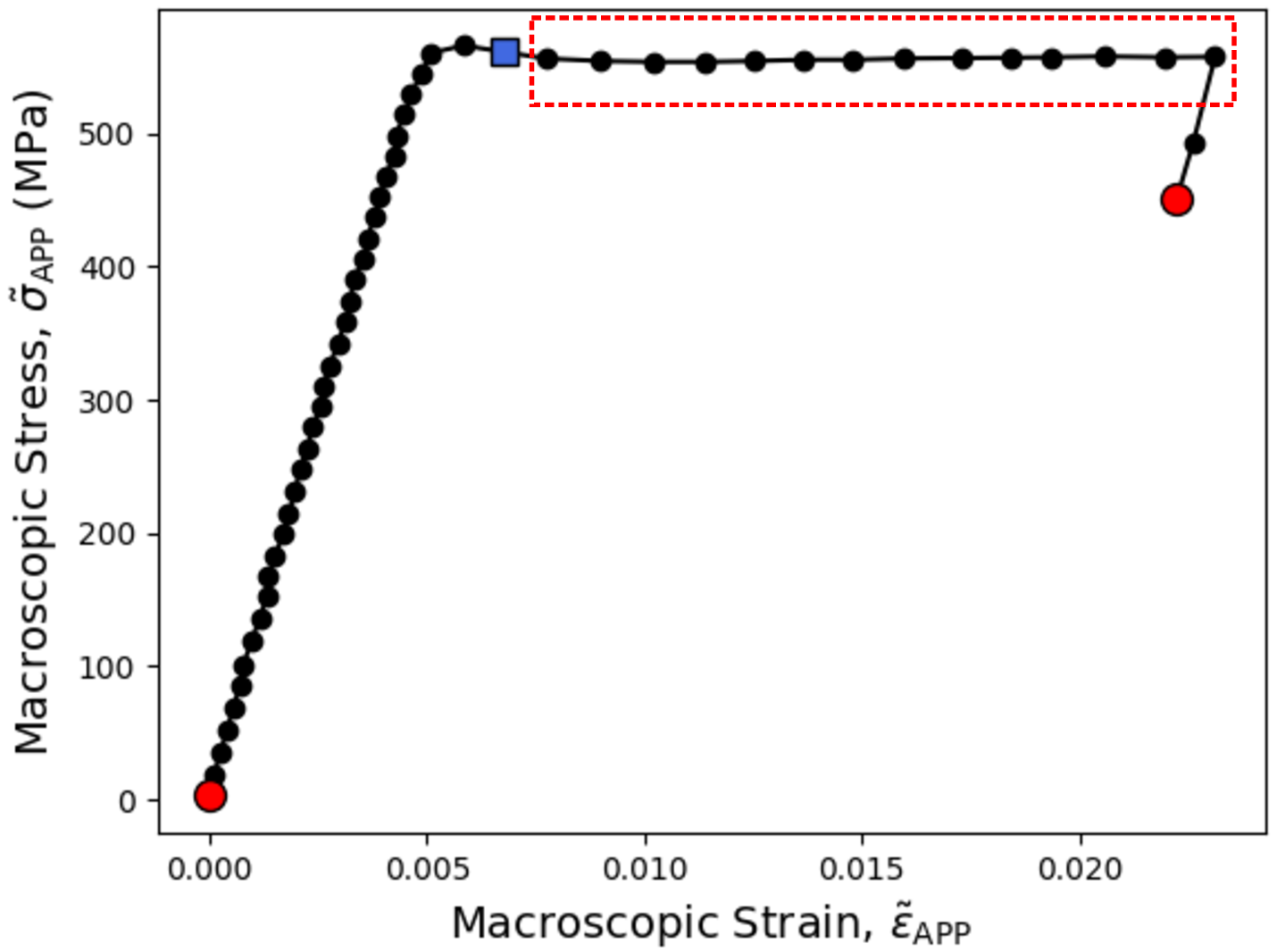}
	  \caption{Macroscopic response of the Ti-7Al specimen ($\tilde{\varepsilon}_{\mathrm{APP}}$ versus $\tilde{\sigma}_{\mathrm{APP}}$). Each black circle corresponds to an ff-HEDM measurement. Red circles indicate the points where loading was halted and nf-HEDM measurements were performed. The blue square indicates the ff-HEDM measurement used to estimate slip system strengths. The outlined red box indicates ff-HEDM measurements used to generate candidate slip systems for each grain.}
	  \label{fig:stressstrain}
\end{figure}

The specimen was deformed in displacement control at a rate of 10$^{-5}$ mm/s to a final applied strain of 2.4\%. The applied macroscopic strain $\tilde{\varepsilon}_{\mathrm{APP}}$ was measured using digital image correlation performed on the the specimen surface and macroscopic stress $\tilde{\sigma}_{\mathrm{APP}}$ was measured using a load cell placed above the specimen and the cross sectional area. The macroscopic response is given in Fig. \ref{fig:stressstrain}. The specimen was characterized by a combination of X-ray  microcomputed tomography ($\mu$CT), nf-HEDM, and ff-HEDM before and during loading. Both the nf-HEDM and $\mu$CT measurements were collected on a LuAG:Ce scintillator coupled to Retiga 4000 DC optical camera with a 5$\times$ lens. Images had 2048 pixels $\times$ 2048 pixels with an effective pixel size of 1.48 $\mu$m. The distance between the scintillator and the specimen was 6.4 mm. The ff-HEDM data were collected on a pair of Dexela 2923 area detectors sitting 884 mm behind the specimen. Each detector had 3888 $\times$ 3072 pixels with a pixel size of 74.8 $\mu$m.

\subsection{HEDM Data Description}
\label{sec:hedm}

\begin{figure}[h]
	\centering
	\includegraphics[width=1.0\textwidth]{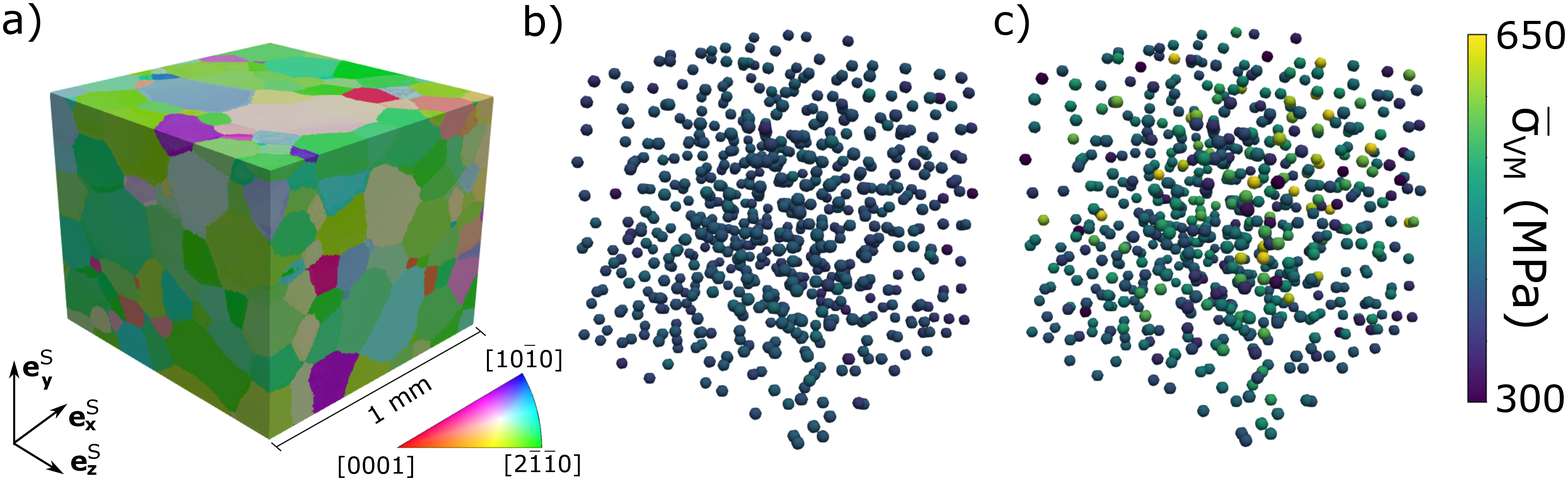}
	  \caption{a) Spatial distribution of lattice orientation in the Ti-7Al volume studied measured using nf-HEDM. The distribution of orientation is colored by orientation of the loading axis with respect to the local lattice orientation. b) Grain-averaged von Mises $\bar{\sigma}_{\mathrm{VM}}$ stresses measured using ff-HEDM directly after yield (indicated with a blue square on Fig. \ref{fig:stressstrain}). c) Grain-averaged von Mises stresses $\bar{\sigma}_{\mathrm{VM}}$ measured using ff-HEDM at a macroscopic applied strain of 2.4\%.}
	  \label{fig:hedm}
\end{figure}

Points where various X-ray measurements were collected are also noted on Fig. \ref{fig:stressstrain}. Red circles indicate points where nf-HEDM data were collected before and after application of uniaxial strain. We note that the specimen was still under 80\% of the maximum applied load when the final nf-HEDM measurements were made to minimize any changes to measured lattice orientation fields due to relaxation. An initial $\mu$CT scan was performed in conjunction to the first nf-HEDM measurements to establish the boundaries of the specimen and to aid the near-field reconstruction. The two nf-HEDM scans were used to measure the spatial distribution of lattice orientation in a 0.75 mm tall volume of the specimen. The full volume was reconstructed from a series of 5 scans using a 0.15 mm tall X-ray beam for each. Fig. \ref{fig:hedm}a shows the reconstructed lattice orientation field for this volume after application of 2.4\% strain. The voxel spacing which defines resolution of the reconstruction is 5 $\mu$m. The reconstruction is colored according to the orientation of the lattice with respect to the loading direction. Inside the near-field volume 592 grains were reconstructed. The algorithm for reconstructing the lattice orientation fields is described in detail in \cite{nygren2020algorithm}. The angular resolution for the reconstructed lattice orientation fields is 0.2$^\circ$. The specimen was sufficiently annealed such that that the initial variation of orientation across the volume probed was less than the angular resolution of the reconstruction. We note that due to the processing of the Ti-7Al, the specimen was textured and as a consequence, relatively few grains with the $<$c$>$ axis of the HCP structure aligned with the loading direction (grains colored red) are observed in Fig. \ref{fig:hedm}a. This texture will influence the slip activity we observed in the specimen.

A series of ff-HEDM scans were collected as the specimen was continuously loaded to measure grain-averaged elastic strain tensors $\bar{\bm{\varepsilon}}_E$, and stresses $\bar{\bm{\sigma}}$, within the Ti-7Al specimen. The algorithm for extracting elastic strain tensors is provided in \cite{bernier2011far}. A single 1 mm tall volume, containing the same grains as those characterized using nf-HEDM, was illuminated for these measurements. The grain data from both the nf-HEDM and ff-HEDM were registered in the process described in \cite{nygren2020algorithm}. However, for the purposes of reconstructing the slip field in the specimen, only ff-HEDM measurements after yield were used. As will be described in the next section, the ff-HEDM measurements collected at this point (marked with a blue square in Fig. \ref{fig:stressstrain}) were used to estimate slip system strengths for prismatic $<$a$>$ (PRI), basal $<$a$>$ (BAS), and pyramidal $<$c+a$>$ (PYR) slip systems. For brevity these families of slip systems will be referred to as prismatic, basal, and pyramidal slip systems without references to Burgers vector. The subsequent ff-HEDM scans were used to build candidate active slip systems for each grain from resolved shear stresses on all possible slip systems.

The grain-averaged stresses were calculated from measured elastic strains from each grain using the aniostropic form of Hooke's Law
\begin{equation}
\bar{\bm{\sigma}}=\bm{C}:\bar{\bm{\varepsilon}}_E.
\label{eq:hooke}
\end{equation}
where $\bm{C}$ is fourth order stiffness tensor. The single crystal elastic moduli used (Voigt notation) are $C_{11}$=162.4 GPa, $C_{12}$=92.0 GPa, $C_{13}$=69.0 GPa, $C_{33}$=180.7 GPa, and $C_{44}$=46.7 GPa \cite{Fisher1964}. Fig. \ref{fig:hedm}b shows the grain-averaged von Mises equivalent stress $\bar{\sigma}_{\mathrm{VM}}$ plotted at each grain's centroid position immediately following yield.  Similarly, Fig. \ref{fig:hedm}c shows the grain-averaged von Mises stress at 2.4\% strain. The average von Mises equivalent stress in each grain is defined as
\begin{equation}
\bar{\sigma}_{\mathrm{VM}}=\sqrt{\frac{3}{2}\bar{\bm{\sigma'}}:\bar{\bm{\sigma'}}}
\label{eq:vm}
\end{equation}
where $\bar{\bm{\sigma'}}$ is the deviatoric portion of the grain-averaged stress.

\section{Slip Reconstruction Algorithm}
In this section, we describe a method for combining spatial distributions of lattice rotation (from nf-HEDM data) and grain-averaged stresses (from ff-HEDM data) to extract 3D spatial fields of slip. As described in the Introduction, characterization of slip is generally performed through analysis of slip by-products with slip inferred through a chosen model. In this paper, the ability to reconstruct slip is made possible through interpretation of the observed lattice rotation through the lens of crystal plasticity. The section is divided in three subsections: kinematics relating measured lattice rotation to slip activity, a method for selecting candidate slip systems, and details regarding how slip activity is calculated across a volume.

\subsection{Reconstruction Kinematics}

\begin{figure}[h]
	\centering
	\includegraphics[width=0.8\textwidth]{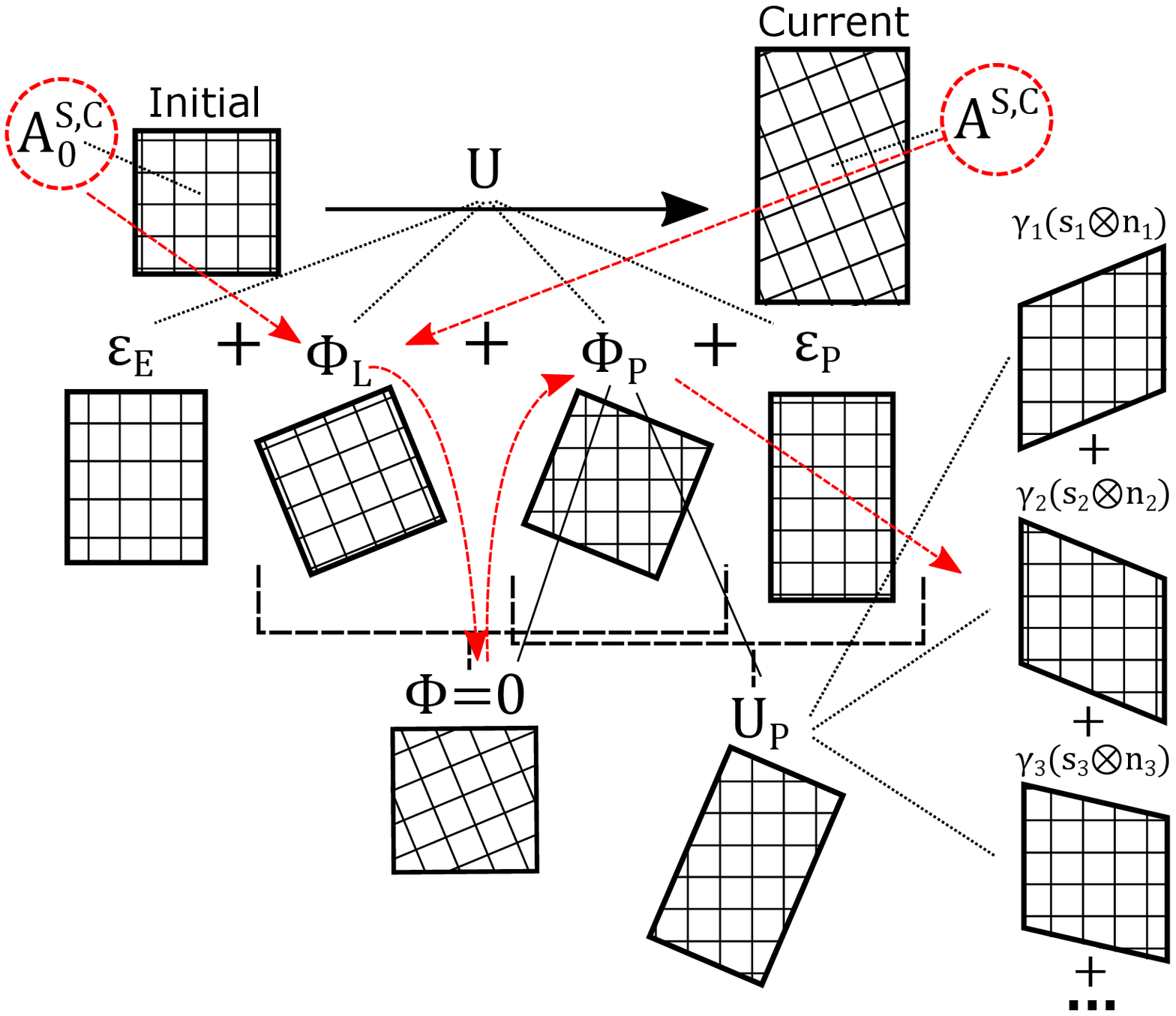}
	  \caption{Schematic of the crystal-scale constitutive assumptions and resulting conditions that form the basis of the presented approach. The crystal lattice orientation during initial and current measurements are represented by $A_0^{S,C}$  and  $A^{S,C}$, respectively. An infinitesimal strain assumption allows for an additive decomposition of $\bm{U}$ into strain and rotation, then continuum slip/crystal plasticity theory is used to further decompose each of those. We assume the continuum rotation, $\bm{\Phi}$, to be zero for each material volume in the unaxially loaded tensile sample. We can then equate $\bm{\Phi_\mathrm{P}}$ to $-\bm{\Phi_\mathrm{L}}$. The plastic distortion of each crystal can then be understood by using the grain-averaged stress to estimate the shear on up to three independent slip systems.}
	  \label{fig:kinematics}
\end{figure}

To aid visualization of the kinematics described in this section, Fig. \ref{fig:kinematics} illustrates the various kinematic quantities incorporated into our constitutive description of the material deformation. The red-arrows indicate the steps by which changes of measured lattice orientation are connected to shear strains on specific slip systems. For slip reconstruction, we employ a linearized decomposition of the deformation gradient $\bm{F}$ into a distortion of the lattice $\bm{U_\mathrm{L}}$ and a plastic distortion $\bm{U_\mathrm{P}}$
\begin{equation}
\bm{F}=\bm{I} + \bm{U_\mathrm{L}}+\bm{U_\mathrm{P}}
\label{eq:defgradient}
\end{equation}
valid in the small-strain regime. Both of these quantities are further decomposed into symmetric and antisymmetric components interpreted as infinitesimal strains $\bm{\varepsilon}$ and rotations $\bm{\Phi}$ of material points
\begin{equation}
\bm{U_\mathrm{L}}=\bm{\varepsilon_\mathrm{E}}+\bm{\Phi_\mathrm{L}}
\label{eq:eldecomp}
\end{equation}
and
\begin{equation}
\bm{U_\mathrm{P}}=\bm{\varepsilon_\mathrm{P}}+\bm{\Phi_\mathrm{P}}.
\label{eq:pldecomp}
\end{equation}

In this small-strain regime, the plastic distortion is described by a sum of shearing motions on the active slip systems in the material.
\begin{equation}
\bm{U_\mathrm{P}}=\sum_{s=1}^{\mathrm{SS}}{\gamma_s(\bm{s}_s \otimes \bm{n}_s)}
\label{eq:slip}
\end{equation}
where $\bm{s}_s$ is a slip direction, $\bm{n}_s$ is a slip plane normal, $\gamma_s$ is the shear strain on a slip system, and $\bm{s}_s \otimes \bm{n}_s$ is the Schmid Tensor. Following fundamental assumptions of crystal plasticity theory, this plastic distortion will deform material points, while leaving the crystal lattice unaltered. As a consequence, only the portion of the deformation associated with distortion of the lattice is measurable using diffraction techniques such as HEDM.

From the kinematics of uniaxial tension, the continuum spin or rotation is identically $\bm{0}$. We assume here that in the small-strain regime for deforming polycrystals, the total rotation of each material volume is negligible and can also be treated as being equal to $\bm{0}$. The measurable skew portion of the lattice deformation, $\bm{\Phi_\mathrm{L}}$, is then equal and opposite to the skew portion of the plastic distortion
\begin{equation}
\bm{\Phi_\mathrm{L}}=-\bm{\Phi_\mathrm{P}}=-\sum_{s=1}^{\mathrm{SS}}{\gamma_s(\bm{s}_s \otimes \bm{n}_s)_{\mathrm{SKEW}}},
\label{eq:rstar1}
\end{equation}
thereby linking the measurable $\bm{\Phi_\mathrm{L}}$ to the plastic distortion that has occurred. Using Eq. \ref{eq:rstar1}, measured slip directions, and measured slip plane normal orientations, plastic shear strains $\gamma_s$ that have occurred on  different slip systems can be determined from a measured $\bm{\Phi_\mathrm{L}}$. As $\bm{\Phi_\mathrm{L}}$ and $\bm{\Phi_\mathrm{P}}$ are skew tensors with three degrees of freedom, plastic shear strains can be uniquely determined on up to three slip systems. The process of attributing candidate slip systems to plastic shear strains is described in the next subsection. We stress at this point that $\bm{\Phi_\mathrm{L}}$ is a rotation of the lattice and not a lattice orientation with respect to an external coordinate system. To calculate $\bm{\Phi_\mathrm{L}}$, both an initial and final lattice orientation are required. We note that once shear strains are attributed to various slip systems, the plastic distortion $\bm{U_\mathrm{P}}$ (and plastic strain $\bm{\varepsilon_\mathrm{P}}$) can be reconstructed using Eq. \ref{eq:slip}.

\subsection{Candidate Slip System Selection}

A list of candidate active slip systems within each grain must be generated prior to solving Eq. \ref{eq:rstar1}. For the slip reconstruction in this work, we assume that all prismatic (3), basal (3), and pyramidal (12) slip systems can activate, for a total of 18 possible slip systems. To create a candidate list, first estimates of grain-averaged shearing rates $\dot{\bar{\gamma}}_s$ are calculated for each possible slip system using a power law formulation relating the slip system shearing rate to resolved shear stress
\begin{equation}
\dot{\bar{\gamma}}_s=\frac{\bar{\tau}_s}{\tau^*_s}\left|\frac{\bar{\tau}_s}{\tau^*_s}\right|^{\frac{1}{m_s}-1}
\label{eq:shearingrate}
\end{equation}
where $\bar{\tau}_s$ is the grain-averaged resolved shear stress on a given slip system, $\tau^*_s$ is the strength of the corresponding slip system, and $m_s$ is a rate sensitivity exponent. As part of this calculation, the grain-averaged resolved shear stress on each slip system $\bar{\tau}_s$, the strength of each slip system $\tau^*_s$, and the rate exponent $m_s$ must be known or assumed. For the primary reconstruction of this work, all slip systems are assumed to have the same rate exponent of 0.02 following \cite{pagan2017modeling}. A discussion of the effect of varying rate exponents on candidate slip system selection is included in \S \ref{sec:rate}.

Using the grain-averaged stresses determined from HEDM data (see \S \ref{sec:hedm}), the grain-averaged resolved shear stress on each slip system is calculated as
\begin{equation}
\bar{\tau}_s=\bar{\bm{\sigma}}:(\bm{s}_s \otimes \bm{n}_s).
\end{equation}
The grain-averaged stresses used to estimate relative slip activity is the last value of stress for the grain past the elastic-plastic transition which was measured with high confidence. For a small number of grains, the associated diffraction peaks became too deformed to reconstruct the elastic strain state with high confidence. In other words, the last value of stress that was determined with high confidence from ff-HEDM measurements is within the dashed red box in Fig. \ref{fig:stressstrain}. For a majority of the grains, this was the final ff-HEDM measurement at 2.4\% applied strain. In this work, grain-averaged stresses which high confidence are those that the ff-HEDM grain parameter reconstruction included over 70\% of predicted peaks (completeness) and had a solution consistency ($\chi^2$) less than 0.01. 

The next step, estimation of slip system strengths, is more of a challenge. A method developed to estimate slip system strengths using grain-averaged stresses measured with ff-HEDM is employed \cite{pagan2017modeling}. In the method, the average strength for a family of slip systems is estimated from the distribution of maximum resolved shear stress across an ensemble of grains. Slip system strengths were determined from the ensemble of grain-averaged stresses measured directly after the elastic-plastic transition (labeled with a blue square in Fig. \ref{fig:stressstrain}). The slip system strengths found using this method are $\tau^*_{\mathrm{PRI}}$=236 MPa, $\tau^*_{\mathrm{BAS}}$=251 MPa, and $\tau^*_{\mathrm{PYR}}$=266 MPa. In the Ti-7Al, evolution of these slip system strengths is minimal due to the low strain-hardening observed.

After determination of $\dot{\bar{\gamma}}_s$ for each slip system within a grain, the slip systems are ordered according to the estimated average shearing rate (i.e., 1 is the most active, 2 is the second most active, and so on). For grains that are estimated to be deforming by single or double slip, the candidate slip system list for each grain is truncated to one or two slip systems (as opposed to the maximum three) to prevent spurious attribution of slip to nonactive systems. Grains deforming by single or double slip are those in which one or two slip systems are estimated to be significantly more active (i.e., $\dot{\bar{\gamma}}_1/\dot{\bar{\gamma}}_2>X$ or $\dot{\bar{\gamma}}_1/\dot{\bar{\gamma}}_3>X$, where $X$ is a large value). For the reconstruction presented, $X$ is set to 100. Again, the shearing rates calculated here are only estimates to enable candidate slip systems to be selected. The determination of the actual amount of slip allocated to each slip system is described in the next section. Lastly, addition of prismatic slip systems as candidate slip systems necessitates extra considerations that are described in more detail in Appendix A.

\subsection{Slip Solution}

To calculate the slip activity at each volume through the reconstruction, first the lattice rotation $\bm{\Phi_\mathrm{L}}$ must be determined. The lattice rotation between an initial and current measurement can be calculated from the components of the lattice orientation matrix $[A^{S,C}]$ (see Fig. \ref{fig:kinematics}). The matrix $[A^{S,C}]$ transforms components of vectors or tensors from a crystal frame to the sample frame. From the nf-HEDM reconstruction, a value of $[A^{S,C}]$ exists for every voxel in the initial ($[A^{S,C}_0]$) and current reconstruction, and it is from these values that $\bm{\Phi_\mathrm{L}}$ is calculated. The components of $\bm{\Phi_\mathrm{L}}$, expressed in the sample coordinate system, are calculated as:
\begin{equation}
(\Phi_L)^S_{ij}=(A^{S,C})_{ik}(A^{S,C}_0)_{jk}-\delta_{ij}.
\label{eq:phi_solve}
\end{equation}
In the case that the initial spatial orientation distribution is constant across grains (or variation is below the measurement resolution), Eq. \ref{eq:phi_solve} can be simplified to 
\begin{equation}
(\Phi_L)^S_{ij}=(A^{S,C})_{ik}(\bar{A}^{S,C}_0)_{jk}-\delta_{ij}.
\label{eq:phi_solve_simple}
\end{equation}
Eq. \ref{eq:phi_solve} or Eq. \ref{eq:phi_solve_simple} is then solved at every point in the current near-field orientation reconstruction. Note that care must be taken registering spatial points between the current and initial configuration; for this work Eq. \ref{eq:phi_solve_simple} is utilized.

With $\bm{\Phi}_L$ determined for each point in the reconstruction, next $\gamma_s$ can be attributed to various slip systems. Eq. \ref{eq:rstar1} is reformulated as a least squares minimization to
\begin{equation}
F(\gamma_s) = \left((\bm{\Phi_\mathrm{L}}+\sum_{s=1}^{\mathrm{CSS}}{\gamma_s(\bm{s}_s \otimes \bm{n}_s)_{\mathrm{SKEW}}}):(\bm{\Phi_\mathrm{L}}+\sum_{s=1}^{\mathrm{CSS}}{\gamma_s(\bm{s}_s \otimes \bm{n}_s)_{\mathrm{SKEW}}})\right)^2
\label{eq:min}
\end{equation}
where the CSS indicates that the summation is performed over a grain's candidate slip systems that are determined previously. In a similar fashion to Eq. \ref{eq:phi_solve_simple}, Eq. \ref{eq:min} is minimized for every point in the reconstruction volume. We note that as both Eq. \ref{eq:phi_solve_simple} and \ref{eq:min} are solved independently for each voxel, calculations are readily parallelized for a reduction in computation time.

\section{Results}
Following the procedure described in the previous section, the nf- and ff-HEDM data collected during elastic-plastic deformation of the Ti-7Al tensile specimen were used to reconstruct the distribution of slip within the probed volume of the specimen. Throughout this section, we will present the distribution of slip throughout the specimen and explore various microstructural and micromechanical effects driving the observed distribution. To facilitate comparison of activity of total slip and slip on various families of slip systems, we introduce the value, $\Gamma$, which is the sum of the absolute value of accumulated slip on different slip systems
\begin{equation}
\Gamma = \sum_{s=1}^{\mathrm{CSS}}{\gamma_s}.
\label{eq:sum}
\end{equation}
Subscripts of this value will denote whether this is the sum of slip activity on all $\Gamma$, prismatic $\Gamma_{\mathrm{PRI}}$, basal $\Gamma_{\mathrm{BAS}}$, or pyramidal $\Gamma_{\mathrm{PYR}}$ slip systems. In the reconstruction, 495 grains (84\%) exhibited prismatic slip, 179 grains (30\%) exhibited basal slip, and 13 grains (2\%) exhibited pyramidal slip. The relatively low number of grains exhibiting pyramidal slip is consistent with the texture of the material. As can be seen in Fig. \ref{fig:hedm}a, there are relatively few grains with the $<$c$>$ axis ([0001] direction) aligned with the loading direction; it is these grains that would be most favorable for pyramidal slip.

\subsection{Distributions of Slip Activity}
\begin{figure}[h]
	\centering
	\includegraphics[width=0.8\textwidth]{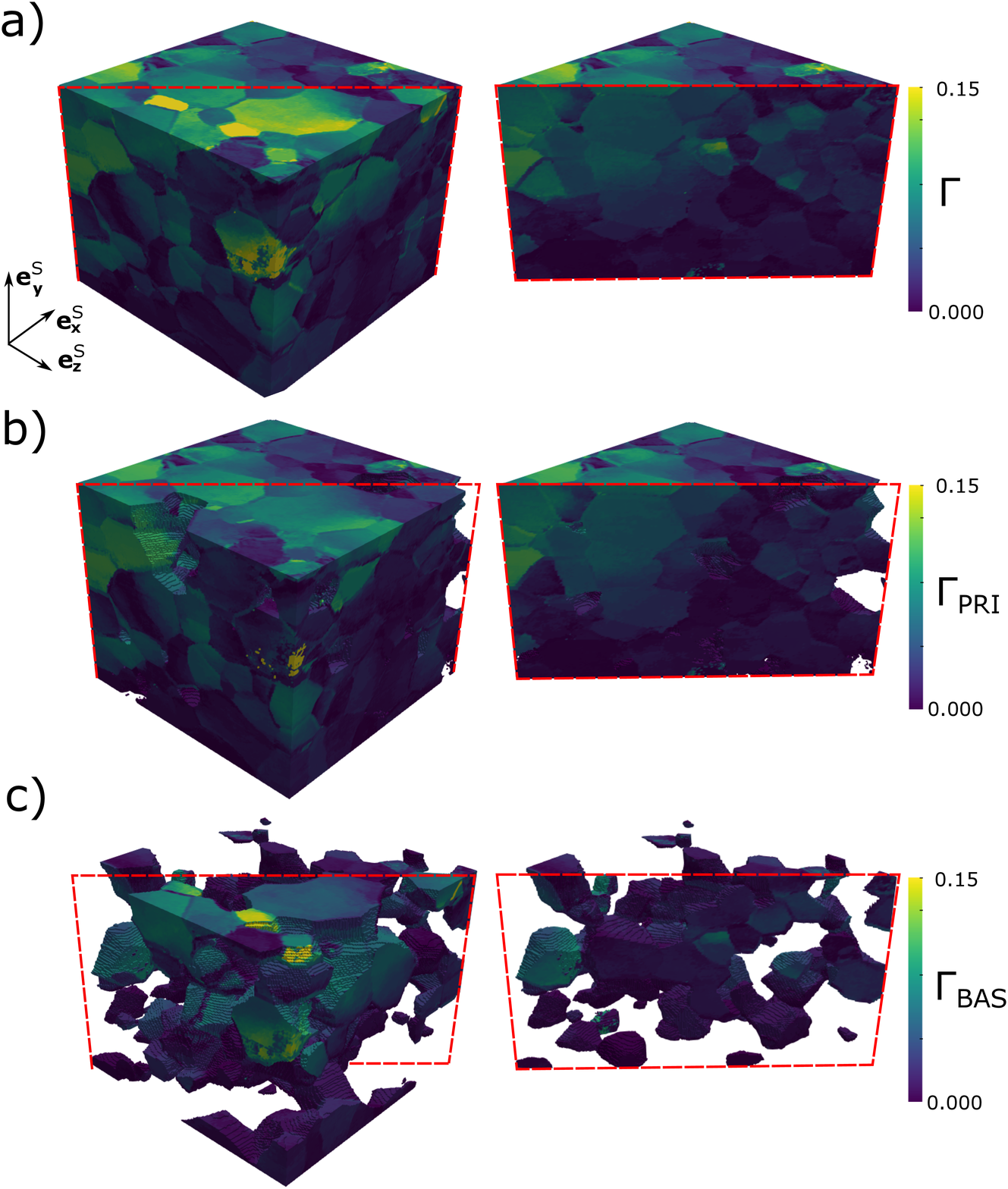}
	  \caption{Intragranular spatial distribution of (including through a cross-sectional cut) of a) total slip $\Gamma$, b) prismatic slip $\Gamma_{\mathrm{PRI}}$, and c) basal slip $\Gamma_{\mathrm{BAS}}$ across the reconstructed Ti-7Al volume.}
	  \label{fig:example}
\end{figure}

Fig. \ref{fig:example} shows the reconstructed distribution of total slip $\Gamma$ (Fig. \ref{fig:example}a) along with distributions of slip for the prismatic $\Gamma_{\mathrm{PRI}}$ (Fig. \ref{fig:example}a) and basal $\Gamma_{\mathrm{BAS}}$ (Fig. \ref{fig:example}b) families of slip systems. Only grains exhibiting prismatic and basal slip are shown in Fig. \ref{fig:example}b and \ref{fig:example}c respectively. The range of total slip increases to values as high as six to seven times the applied macroscopic strain $\tilde{\varepsilon}_{\mathrm{APP}}$ (0.024), however most of the volume displays significantly less total shear strain. A cross section of each slip distribution is also shown (outline denoted with dashed red lines). We note that the regions around the surface of the specimens are those with the lowest reconstruction confidence (ratio of predicted versus measured intensity) in the nf-HEDM orientation reconstruction which may explain the small regions containing what appears to be spurious elevated shear strain levels around the boundary.

\begin{figure}[h]
	\centering
	\includegraphics[width=1.0\textwidth]{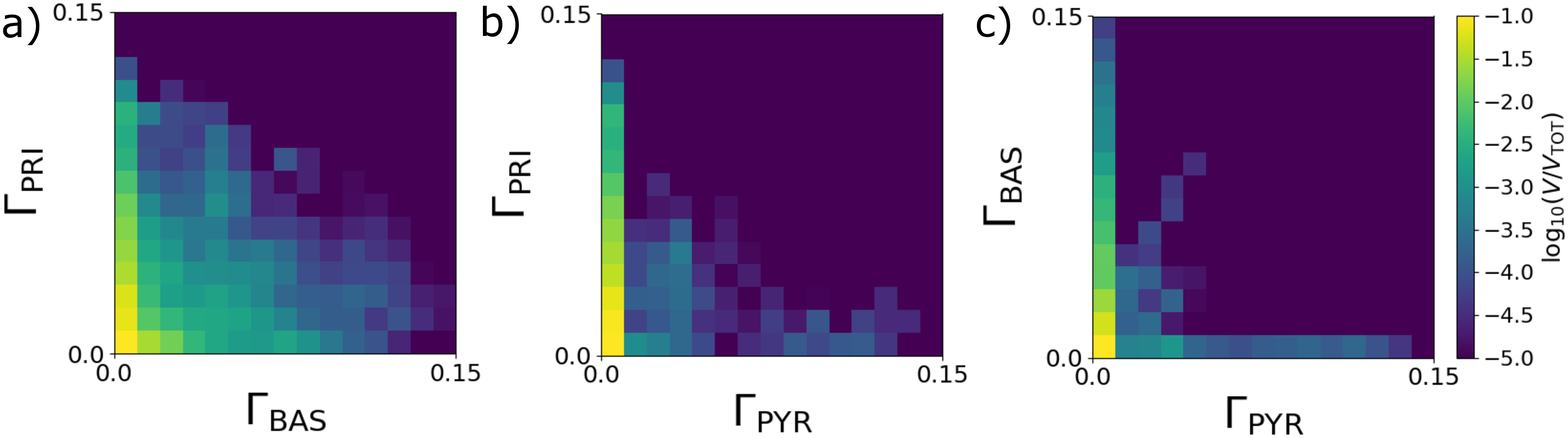}
	  \caption{Two-dimensional histograms showing volume fractions ($V/V_{\mathrm{TOT}}$) exhibiting different combinations of slip activity within the reconstructed volume, including a) prismatic and basal slip, b) prismatic and pyramidal slip, and c) basal and pyramidal slip. Volume fractions are shown on a logarithmic scale. }
	  \label{fig:slipact}
\end{figure}

Fig. \ref{fig:slipact} shows the relative volume fractions ($V/V_{\mathrm{TOT}}$) of different combinations of slip activity throughout the reconstructed slip field on a series of 2D histograms. The volume fractions are calculated by dividing the number of voxels with a given slip activity by the total number of voxels and are plotted on a logarithmic scale (a value of -1 corresponds to 10\% of the reconstructed volume and -5 corresponds to 0.001\%). Fig. \ref{fig:slipact}a shows prismatic and basal activity, Fig. \ref{fig:slipact}b shows prismatic and pyramidal activity, and Fig. \ref{fig:slipact}c show basal and pyramidal activity. From the figures, we can see that slip activity dominated by a single slip system is most common at the relatively low applied strain (high volume fractions on the left and bottom edges of the histograms). Fig. \ref{fig:slipact}a also shows that a large volume fraction of material exhibits both prismatic and basal slip simultaneously. Lastly, Fig. \ref{fig:slipact}b and \ref{fig:slipact}c indicate that pyramidal slip in combination with either pyramidal or basal slip is relatively rare, but prismatic and pyramidal occurring together is more prevalent.

\subsection{Orientation Dependence of Slip Activity}
\label{sec:oridep}

\begin{figure}[h]
	\centering
	\includegraphics[width=1.0\textwidth]{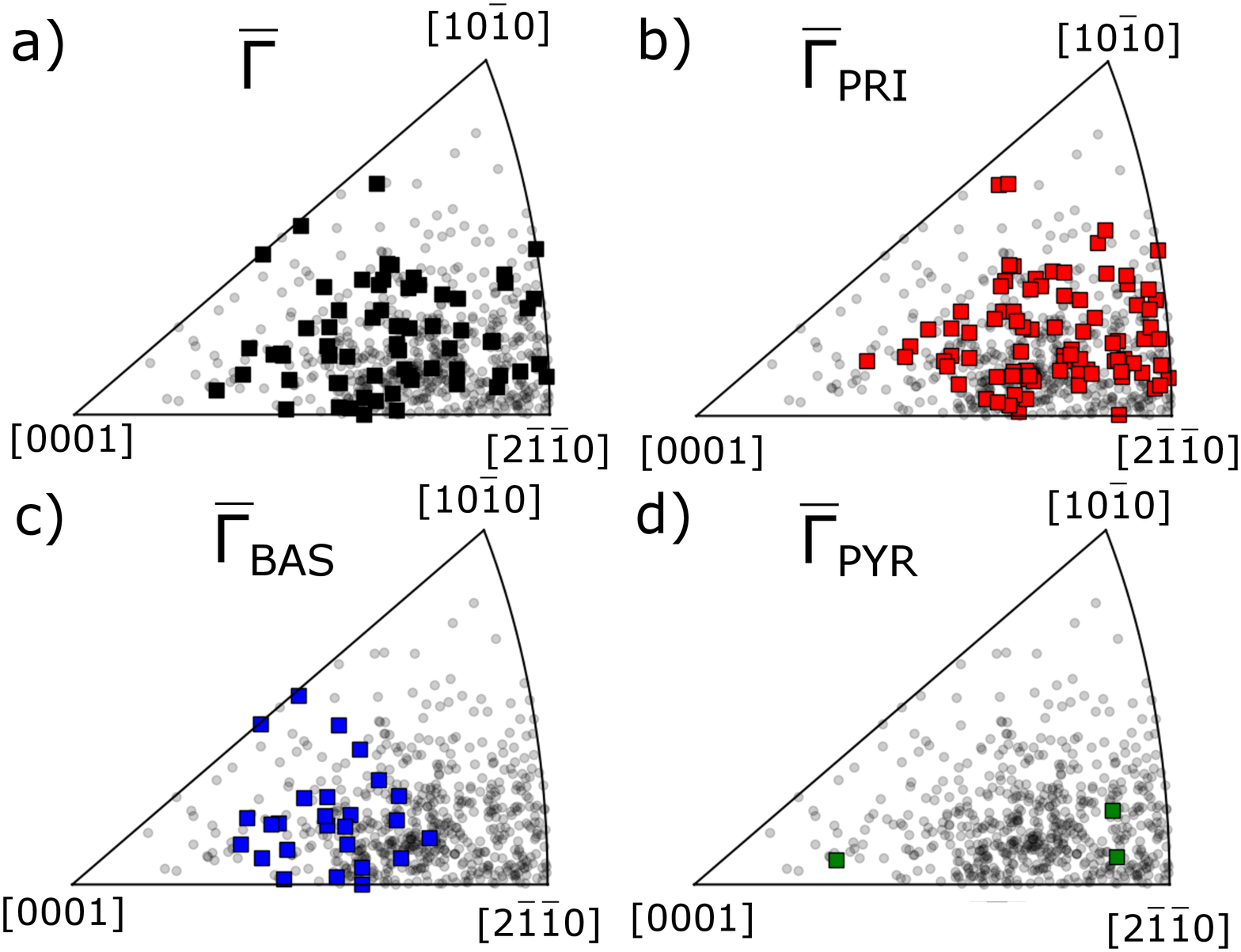}
	  \caption{Orientations of grains exhibiting high values of grain-averaged slip plotted on the hexagonal fundamental triangle with respect to the loading axis. a) Orientations of grains (black squares) with total slip more than 3$\times$ the macroscopic applied plastic strain. b) Orientations of grains (red squares) with prismatic slip more than 2$\times$ the macroscopic applied plastic strain.  c) Orientations of grains (blue squares) with basal slip more than 2$\times$ the macroscopic applied plastic strain. d) Orientations of grains (green squares) with pyramidal slip more than 2$\times$ the macroscopic applied plastic strain.}
	  \label{fig:oridep}
\end{figure}

An analysis of the orientation dependence of grains exhibiting various types of slip was performed to confirm the fidelity of the slip reconstruction. Grains exhibiting relatively large amounts of slip were extracted from the slip reconstruction and then plotted according to their orientation with respect to the loading axis on the fundamental triangle for hexagonal crystals. Grains with large amounts of slip were determined by comparing the grain-averaged accumulated slip magnitudes to the applied plastic strain (0.019). These include grains with accumulated amounts of slip greater than 3$\times$ the applied plastic strain and grains with slip on a particular family of slip systems greater than 2$\times$ the applied plastic strain. Fig. \ref{fig:oridep}a shows the grains with high total amount of slip $\bar{\Gamma}$ on the hexagonal fundamental triangle while Fig. \ref{fig:oridep}b, \ref{fig:oridep}c, and \ref{fig:oridep}d show the grains exhibiting large amount of prismatic $\bar{\Gamma}_{\mathrm{PRI}}$, basal $\bar{\Gamma}_{\mathrm{BAS}}$, and pyramidal $\bar{\Gamma}_{\mathrm{PYR}}$ slip respectively. Other orientations with low amounts of slip in the reconstruction volume are plotted with reduced opacity on each fundamental triangle. In Fig. \ref{fig:oridep}a, we can see that there is little orientation dependence of grains exhibiting large amounts of slip since the grains are evenly distributed over the fundamental triangle. However, as would be expected, grains that exhibit large amounts of prismatic and basal slip do have an orientation dependence. The grains exhibiting a large amount of prismatic slip tend to be closer to the [$10\bar{1}0$]-[$2\bar{1}\bar{1}0$] boundary (Fig. \ref{fig:oridep}b), and the grains exhibiting large amounts of basal slip tend to reside in the center of the fundamental triangle (Fig. \ref{fig:oridep}c). As stated previously, there are relatively few grains exhibiting any pyramidal slip in the reconstruction and, as such, there similarly are not many grains showing large amounts of pyramidal slip. 

\subsection{Stress Dependence of Slip Activity}

\begin{figure}[h]
	\centering
	\includegraphics[width=1.0\textwidth]{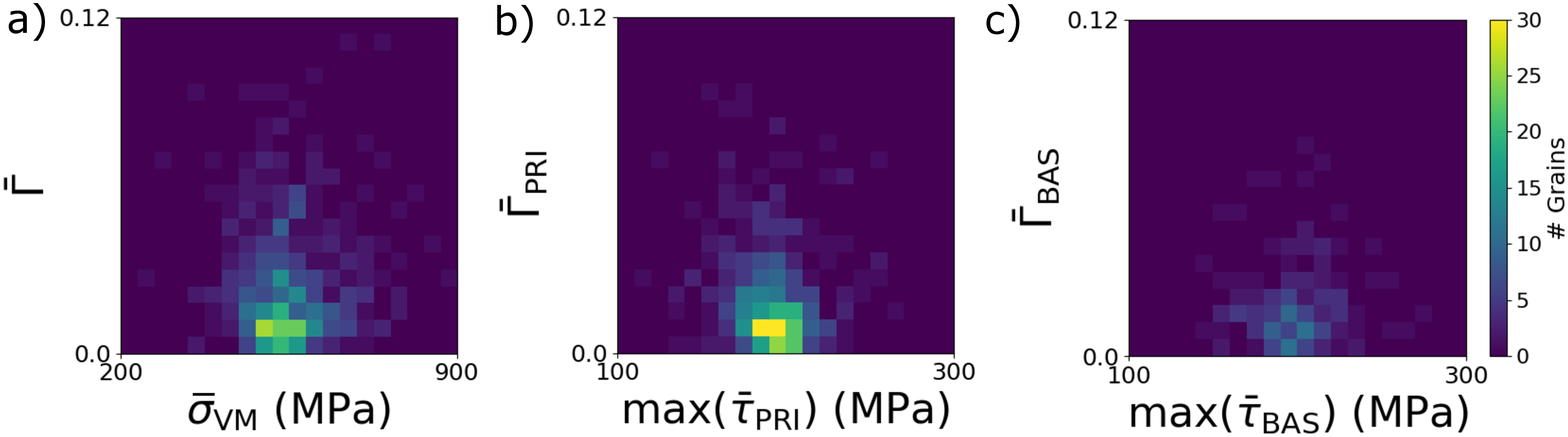}
	  \caption{Two-dimensional histograms of grain-averaged a) von Mises stress versus total slip, b) maximum prismatic resolved shear stress versus prismatic slip, and c) maximum basal resolved shear stress versus basal slip.}
	  \label{fig:stressdep}
\end{figure}

The full local stress state and the shear stresses applied to slip systems are primary drivers of slip activity within a polycrystal. However, the total amount of observed slip is also dependent on constraints of local boundary conditions applied to grains. To explore correlations between stress magnitude and slip activity, 2D histograms were generated by binning grains according to grain-averaged total slip $\bar{\Gamma}$ and von Mises stress $\bar{\sigma}_{\mathrm{VM}}$; prismatic slip $\bar{\Gamma}_{\mathrm{PRI}}$ and maximum slip on a prismatic slip system max($\bar{\tau}_{\mathrm{PRI}}$); and basal slip $\bar{\Gamma}_{\mathrm{BAS}}$ and maximum slip on a basal slip system max($\bar{\tau}_{\mathrm{BAS}}$). These three histograms can be seen in Fig. \ref{fig:stressdep}a, \ref{fig:stressdep}b, and \ref{fig:stressdep}c respectively. While not directly tied to slip activity, $\bar{\sigma}_{\mathrm{VM}}$ provides a convenient scalar measure of stress to explore broad trends. In the figures, we find relatively little correlation between stress magnitude and the average slip activity within a grain. If there was a positive correlation, the distributions would exhibit a positive slope. Instead the distributions extend primarily vertically for the three distributions shown. This indicates that while stress is important for activating slip systems, the magnitude of slip may instead be dictated by the need to accommodate both macroscopic and local boundary conditions. As a reminder, the reconstruction algorithm presented only considers stress in the selection of candidate slip systems, the magnitude of slip is determined from the lattice rotation field.

\subsection{Connectivity of High Slip Activity}

\begin{figure}[h]
	\centering
	\includegraphics[width=1.0\textwidth]{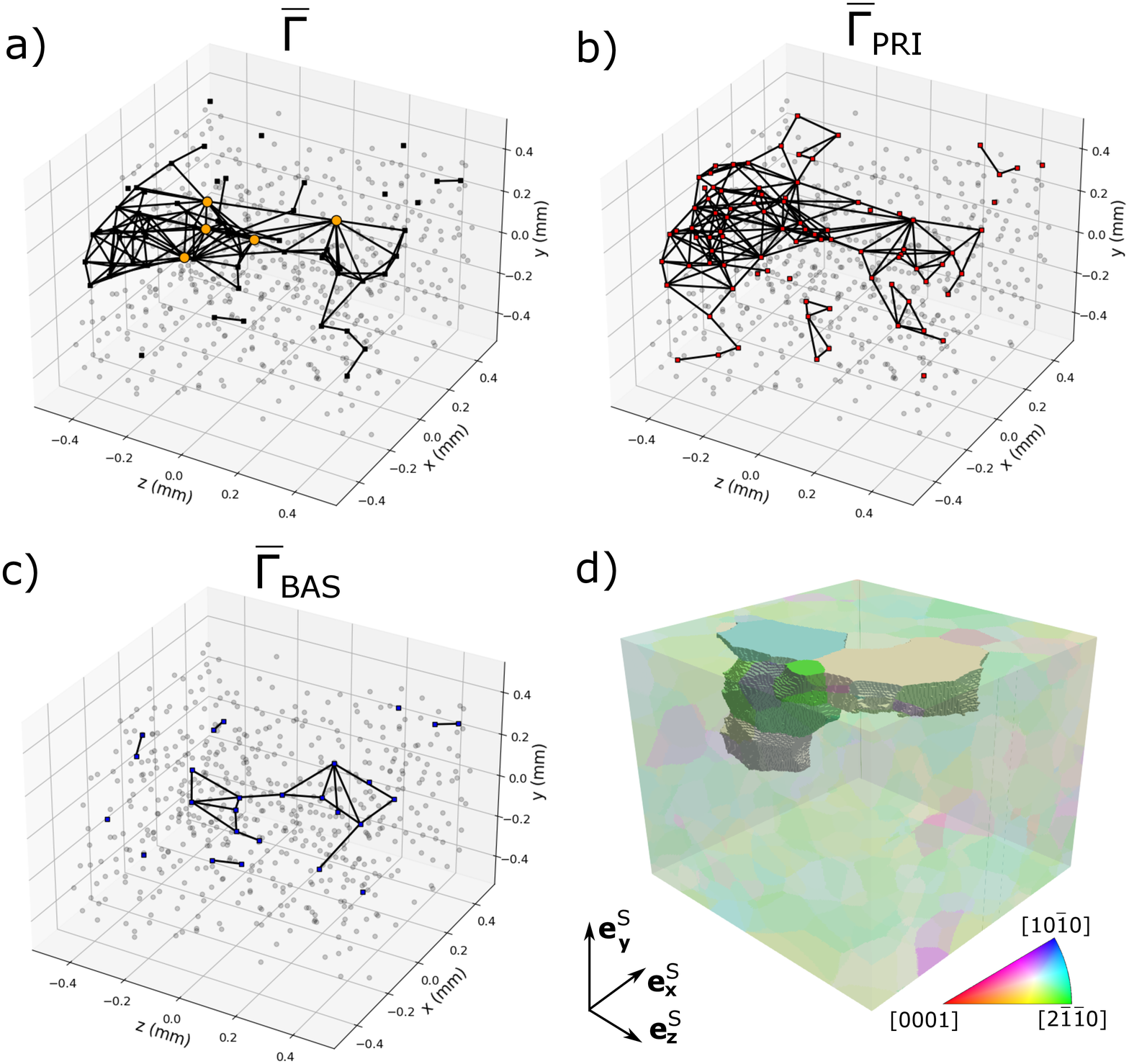}
	  \caption{a-c) Centroids determined from ff-HEDM data of grains exhibiting large amount of slip. Shared boundaries between high slip grains are marked with black lines. a) Centroids of grains (black squares) with grain-averaged total slip $\bar{\Gamma}$ more than 3$\times$ the macroscopic applied plastic strain. Orange circles denote the five grains with the highest betweenness centrality. b) Centroids of grains (red squares) with grain-averaged prismatic slip $\bar{\Gamma}_{\mathrm{PRI}}$ more than 2$\times$ the macroscopic applied plastic strain.  c) Centroids of grains (blue squares) with grain-averaged basal slip $\bar{\Gamma}_{\mathrm{BAS}}$ more than 2$\times$ the macroscopic applied plastic strain. d) The morphologies and orientations determined from nf-HEDM data of the five grains within the extended network of high slip grains that have the largest betweenness centrality.}
	  \label{fig:networks}
\end{figure}

While orientation and stress resolved on to slip systems play a major role in activation of slip systems, these quantities do not seem to dictate the magnitude of slip observed in embedded grains. To explore spatial dependence of slip activity, grains that exhibited elevated amounts of slip as defined in \S \ref{sec:oridep} are plotted with respect to their centroids in 3D. Fig. \ref{fig:networks}a, \ref{fig:networks}b, and \ref{fig:networks}c show `high slip grains' exhibiting high total slip, prism slip, and basal slip respectively. In Fig. \ref{fig:networks}a, \ref{fig:networks}b, and \ref{fig:networks}c grain centroids from the ff-HEDM data are marked with squares and connections (shared grain boundaries) between high slip grains are marked with black lines. What is clear from these figures is that a large fraction of high slip grains exist in a connected network. Some isolated high slip grains and clusters of grains with 2 to 5 grains participating exist, but most high slip grains are located in a centralized grouping of connected grains. These results suggest that not only are the immediate neighbors critical factors in the amount of slip experienced by a grain, but second and third neighbors also play a critical role as slip is transferred through networks of grains. A deeper discussion of the large group of connected high slip grains will be provided shortly. Importantly, this insight regarding connectivity of slip transmission could only be gleaned from 3D analysis of slip activity.

\section{Discussion}

The 3D distribution of slip that developed in Ti-7Al during uniaxial tension to 2.4\% strain was reconstructed. The novel reconstruction method presented is made possible by utilizing both nf-HEDM and ff-HEDM data throughout the slip reconstruction process. The fundamentals of continuum slip (crystal plasticity) modeling were used as a framework for the interpretation of the HEDM data in a way that forms the basis of our approach (illustrated schematically in Fig. \ref{fig:kinematics}). In the strain regime analyzed here, an infinitesimal strain assumption is justified, which allows for the additive decomposition of deformation and also allows us to speak of strains and rotations instead of strain rates and spins. The uniaxial deformation of the test specimen (applied velocity along the loading axis) produces a rotation-free condition on the macroscale. We impose that condition on each material volume, enabling us to equate the lattice rotation, $\bm{\Phi_\mathrm{L}}$ – which we are able to monitor in the HEDM experiment -  with the negative of the rotation due to continuum slip plasticity, $\bm{\Phi_\mathrm{P}}$. While each volume of material may not be experiencing pure uniaxial extension, this deformation mode will surely dominate the kinematics of each crystal – especially in this strain regime. Now with an estimate of the rotation due to continnum slip, we use grain-averaged stresses to propose which slip systems (up to three) are active to produce that shear. 

In the reconstruction process, grain-averaged stresses and estimates of slip system strengths are directly determined from the ff-HEDM data, as opposed to being assumed from Schmid factor analysis, and then used to build lists of candidate active slip systems in each grain. Lattice rotation fields are then determined from evolving nf-HEDM data and are used to reconstruct the magnitude of slip on different slip systems through the polycrystal. The ability to reconstruct distributions of slip non-destructively and in 3D through the bulk of a polycrystal is a major advance in slip analysis efforts that have extended over 80 years. We emphasize that while a single plastic strain increment is reconstructed in this work, the X-ray diffraction measurements were performed in-situ at load. Thus, the method can be used to quantify slip at multiple points during deformation without having to fully unload the specimen, enabling the evolution of 3D slip distributions to be quantified.

 Pioneering work in fatigued single crystals identified the importance of slip transmission in the fatigue life of deformed single crystals \cite{cheng1981fatigue,finney1975strain,witmer1987nucleation}. In the case of ductile single crystals under fatigue loading, bands along which slip preferentially occurred were associated with the formation of fatigue cracks at band termination points on the sample surface. While analogous slip structures likely exist in deforming polycrystals, the ability to identify them has had limited success, possibly due to an inability to monitor slip through grains in a polycrystal as it is deforming. Here, with the new capability presented, we have identified a connected group of grains where slip appears to preferentially travel in the Ti-7Al specimen studied. While significant further study is required, the measurements shown and the capability presented provide a path forward for extending our understanding of slip transmission past single crystals and single grain boundaries to determining the role of slip networks in the mechanical response of deforming polycrystals.

\subsection{Slip Network Analysis}

Due to the connected nature of the large group of grains exhibiting large amounts of slip within the reconstructed volumes (Fig. \ref{fig:networks}a), treating this group of grains as a network \cite{lenthe2020twin} provides a natural framework with which to analyze the slip transmission. In the network framework, each grain is a node of the network while each shared boundary is a connection (also referred to as an edge). Using this framework, we can employ various measures to quantify and rank the structural importance (centrality) of various grains in the slip network \cite{newman2018networks}. The first measure is the \emph{degree centrality} which quantifies the number of connections (normalized by the total number of connections in the graph) directly into a grain. The degree centrality is a relatively simple measure similar to a coordination number. The second measure is the \emph{eigenvector centrality}, which ranks grains by the number and \emph{and importance} of its neighbors. For the purpose of this analysis, grains with high eigenvector centrality transmit slip to neighboring grains that in turn also transmit slip to a relatively large number of grains. The final centrality measure is the \emph{betweenness centrality}. The betweenness centrality differs from the other two measures in that it ranks the importance of nodes (grains) within paths as opposed to the number of connections to the node itself (shared grain boundaries). The betweenness centrality scores nodes according to the number of paths between two points in the network that pass through the node. In this context, the betweenness centrality ranks a grain according to the number of slip transmission paths that pass through it, and as such, likely provides the most insight into importance of grains in slip transmission among the centrality measures presented.

The grains with the five highest scores for each of these three centrality measures within the high slip grouping of grains is given in Table \ref{tab:centab}. Extraction of these quantities was performed with the NetworkX software package \cite{hagberg2008exploring}. Grain numbering is arbitrary.  In the table, the exact numerical values of the centrality measures are less of a focus than which grains have the highest values. In the table, we see that there is significant amount of overlap between the ranked lists of centrality measures. Grain 440 has the highest values for all three measures, underscoring its importance, and grain 299 is in the top three for all measures. The heavy overlap between the listings highlights the often related nature of many centrality measures, but we will focus on the betweenness centrality which, as mentioned, provides insight into grains that can serve as `conduits' for slip transmission through the network. The five grains with the highest betweenness values are shown in Fig. \ref{fig:networks}d. The common characteristics of these grains is that they all appear to be connected, relatively large in size, and oriented favorably for prismatic slip. We note the commonalities between this group of grains that exhibits high betweenness centrality and a `macrozone' of similarly oriented grains favorable for slip often observed in titanium alloys. However, our ability to characterize the 3D slip network around these grains provides new context. Connected groups of large grains with orientations favorable for slip appear to drive large amounts of slip in an extended network of grains well past immediate neighbors (see Fig. \ref{fig:networks}a). Further analysis of these effects and slip propagation with increasing applied strain using the reconstruction method presented in this work may provide significant insight into the extended effects of slip driven by macrozones and their subsequent effects on fatigue in titanium alloys.

\begin{table}
	\centering
	\begin{tabular}{|c|c|c|c|c|c|c|}
	    \hline
		&\multicolumn{2}{c|}{Degree} & \multicolumn{2}{c|}{Eigenvector}& \multicolumn{2}{c|}{Betweenness} \\ \hline
		Rank & No. &  Val. & No. &  Val. & No. &  Val.\\ \hline
		1 & 440 &  0.025 & 440  & 0.373 & 440 & 0.0029\\ \hline
		2 & 423 &  0.019 & 299 & 0.284 & 423 & 0.0028\\ \hline
		3 & 299 &  0.019 & 543 & 0.278 & 299 & 0.0021\\ \hline
		4 & 543 &  0.017 & 310 & 0.271 & 425 & 0.0013\\ \hline
		5 & 425 &  0.017 & 473 & 0.248 & 310 & 0.0009\\ \hline
	\end{tabular}
\caption{Ranking of structural importance (centrality) of the top five grains in the elevated slip network as defined by various centrality measures including degree, eigenvector, and betweenness.}
\label{tab:centab}
\end{table}

\subsection{Effects of Rate Exponent Selection}
\label{sec:rate}
Historical efforts to understand activation of slip in hexagonal Ti alloys has primarily focused on relative slip system strengths (or critical resolved shear stresses) to predict microscale slip behavior, but the effects of varying rate sensitivity across families of slip systems is increasingly acknowledged as playing a critical role. Recent work has presented evidence of differing rate sensitivities for various families of slip systems in hexagonal titanium alloys \cite{jun2016local,zhang2016intrinsic}. In addition, variation of rate sensitivity of different families of slip systems may play a role in dwell fatigue failure in Ti alloys \cite{zhang2015rate}. While the choice here of using a rate sensitivity exponent of 0.02 for all families of slip system (Set 1) has been shown to capture the grain-averaged response of Ti-7Al \cite{pagan2017modeling}, it is worth exploring how variation of the rate sensitivities chosen for the slip reconstruction method will affect the construction of candidate slip system lists and subsequently the slip reconstruction. 

\begin{figure}[h]
	\centering
	\includegraphics[width=0.7\textwidth]{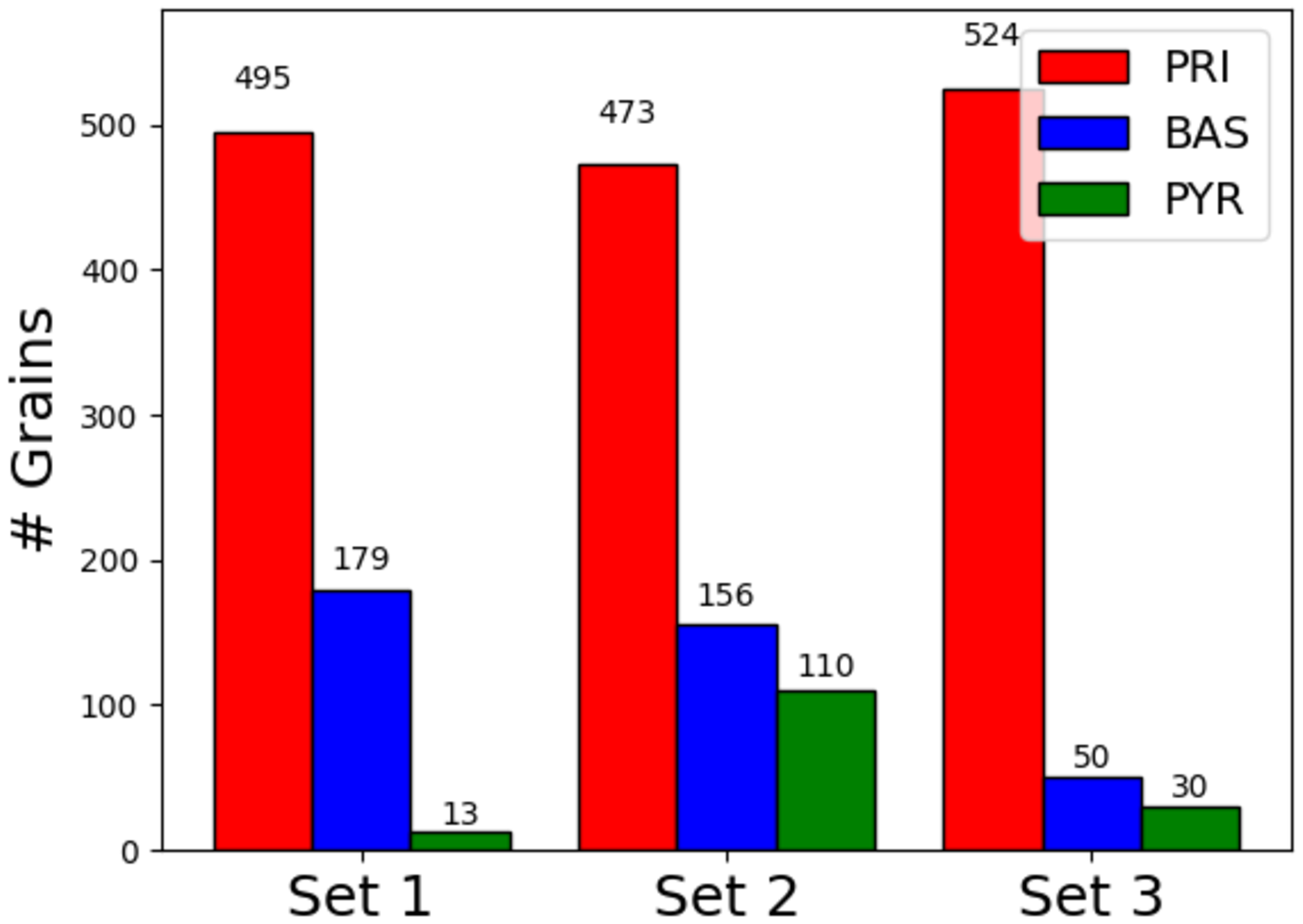}
	  \caption{Histogram showing the number of grains exhibiting slip on different families of slip systems using various sets of rate exponents. The composition of each set is: Set 1 (standard set used in this work): $m_{\mathrm{PRI}}$=0.02, $m_{\mathrm{BAS}}$=0.02, $m_{\mathrm{PYR}}$=0.02; Set 2 : $m_{\mathrm{PRI}}$=0.02, $m_{\mathrm{BAS}}$=0.02, $m_{\mathrm{PYR}}$=0.04; Set 3: $m_{\mathrm{PRI}}$=0.04, $m_{\mathrm{BAS}}$=0.02, $m_{\mathrm{PYR}}$=0.04.  }
	  \label{fig:exponent_effects}
\end{figure}

To determine how different choices of rate sensitivities influence the slip reconstruction, two other sets of rate exponents (Set 2 and Set 3) were selected and used to repeat the slip reconstruction process. Set 2 corresponds to doubling the rate sensitivity of pyramidal slip systems ($m_{\mathrm{PRI}}$=0.02, $m_{\mathrm{BAS}}$=0.02, $m_{\mathrm{PYR}}$=0.04) and Set 3 corresponds to doubling rate sensitivity of both prismatic and pyramidal slip systems ($m_{\mathrm{PRI}}$=0.04, $m_{\mathrm{BAS}}$=0.02, $m_{\mathrm{PYR}}$=0.04). Fig. \ref{fig:exponent_effects} shows a histogram of the number of grains exhibiting different types of slip using each of the various combinations of rate exponents. While changing the rate exponent does not radically change the ratios of slip through the polycrystal (prismatic slip is still dominant), increasing the rate exponent increases the amount of slip observed on a given family of slip systems. For Set 2, increasing the pyramidal rate exponent significantly increases the number of grains that exhibit pyramidal slip, while Set 3 shows increases in both prismatic and pyramidal slip. Also using Set 2 and Set 3, the increases in pyramidal or prismatic slip also is accompanied by reductions of the amount of basal slip found. The reason for this is that as the rate exponent increases, activation of slip becomes a more gradual process, so slip will generally occur at lower resolved shear stress levels, allowing for small amounts of slip on secondary slip systems. These results further point to rate sensitivity playing a critical role in the slip behavior of $\alpha$ Ti alloys, and further efforts attempting to isolate the rate sensitivity of different families of slip within deforming polycrystals is warranted.

%\subsection{Application to Other Crystal Types}

\section{Summary}

A novel method for non-destructively reconstructing 3D slip fields at subgrain length scales using a combination of nf- and ff-HEDM data is presented. The method is applied to reconstructing the distribution of slip in nearly 600 Ti-7Al grains after application of 2.4\% applied uniaxial strain. From analysis of the reconstructed 3D slip field we find that:
\begin{itemize}
    \item For the orientation distribution present within the probed volume, slip is dominated by prismatic slip with some basal slip also present. Minor amounts of pyramidal $<$c+a$>$ slip are also observed, consistent with few grains with the $<$c$>$ axis aligned with the loading direction being present.
    \item As expected, whether a grain is found to have a large amount of prismatic or basal slip is highly dependent on the crystal orientation.
    \item Minimal correlation is found between the magnitude of various grain-averaged stress measures (von Mises, maximum prismatic resolved shear stress, and maximum basal resolved shear stress) and the magnitude of slip within a grain.
    \item Grains exhibiting elevated amounts of slip appear to be part of an extended network of grains that are centered around a small number of relatively large grains that are favorably oriented for (prismatic) slip. These large grains were found to have significant structural importance within the network.
\end{itemize}
This novel capability to reconstruct slip fields as polycrystals are plastically deforming provides future opportunities to explore many inherently 3D features of slip transmission, including whether slip in polycrystals is associated with persistent network structures.

\section{Acknowledgements}
This work is supported financially by the Office of Naval Research, United States Department of Defense, contract N00014-16-1-2982, Dr. William Mullins, program manager.  This work is based upon research conducted at the Center for High Energy X- ray Sciences (CHEXS) which is supported by the National Science Foundation (United States) under award DMR-1829070. KEN is supported by the Materials Solutions Network at CHESS (MSN-C) which is supported by the Air Force Research Laboratory under award FA8650-19-2-5220. The Ti-7Al material was provided by Dr. Adam Pilchak, Air Force Research Laboratory, United States. The authors would like to thank Professor Matthew Kasemer for helpful discussions.

\bibliography{bibliography}

\section*{Appendix A: Considerations for Reconstructions of Prismatic Slip}
\label{sec:prism}
\begin{figure}[h]
	\centering
	\includegraphics[width=0.5\textwidth]{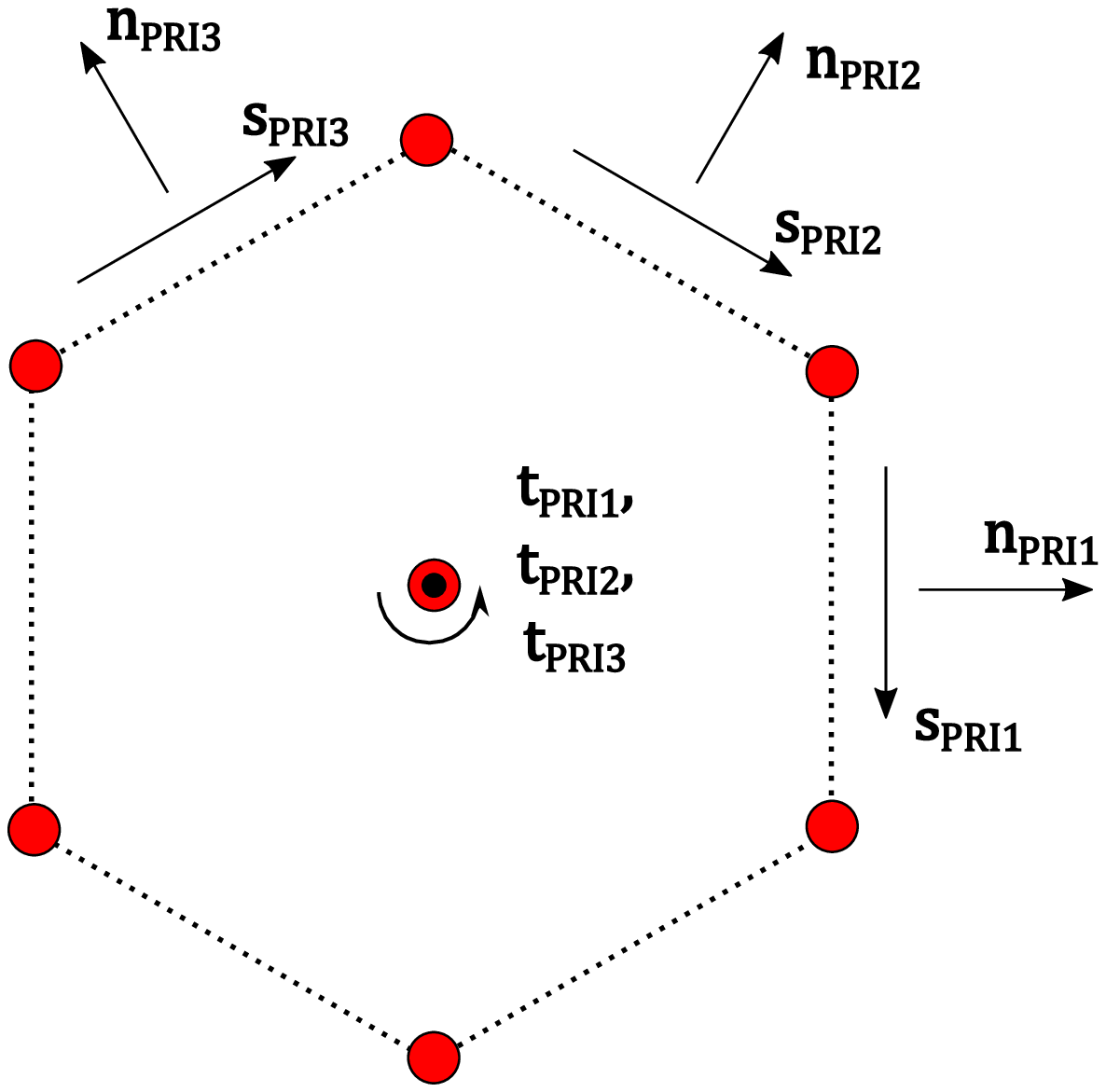}
	  \caption{Schematic of the shared rotation axis $\bm{t}_\mathrm{PRI}$ for the three prismatic slip systems defined by slip direction $\bm{s}_\mathrm{PRI}$ and slip plane normal $\bm{n}_\mathrm{PRI}$.}
	  \label{fig:prism_slip}
\end{figure}

The inclusion of prismatic slip systems into the slip reconstruction algorithm requires extra consideration for the assembly of candidate slip systems. As mentioned, Eq. \ref{eq:rstar1} can be used to determine $\gamma_s$ for three linearly independent slip systems. However, the skew portions of the Schmid Tensors ($\bm{s}_s \otimes \bm{n}_s$) for the three prismatic slip systems are not linearly independent, producing issues for slip reconstruction. One means to understand these issues is to analyze the axes about which individual shearing motions rotate the material. We can reformulate \cite{pagan2016determining} $\bm{\Phi_\mathrm{P}}$ to
\begin{equation}
\bm{\Phi_\mathrm{P}}=-\sum_{s=1}^{\mathrm{SS}}{\frac{\gamma_s}{2}(\bm{t}_s \times \bm{I})}
\end{equation}
where
\begin{equation}
\bm{t}_s=\bm{s}_s \times \bm{n}_s
\end{equation}
and $\times \bm{I}$ indicates the operation of constructing a skew tensor from a vector. Fig. \ref{fig:prism_slip} illustrates the directions $\bm{t}_s$ about which material will rotate during slip on the three prismatic slip systems.  In the figure we can see that because the slip directions and slip planes for the prismatic slip systems are all co-planar, they share the same rotation axes. Due to this redundancy, candidate slip system lists for hexagonal alloys deforming by prismatic slip (e.g., Ti-7Al) can only contain a single prismatic slip system. Here, the prismatic slip system estimated to be most active is selected if two of three prismatic slip systems happen to satisfy the previously described selection conditions. Due to geometric considerations, the case of three prismatic slip system being the highest ranked candidate slip systems is not encountered during uniaxial tension. With the described approach taken for slip reconstruction, the output from the slip reconstruction algorithm is essentially a `net' slip of the prismatic slip systems with the upper bound reported equal to the total slip across all slip systems. We note that this choice only affects a subset of grains in the slip reconstruction.

\end{document}